\documentclass[12pt]{article}
\usepackage{graphicx}
\usepackage{epsfig}
\usepackage{epstopdf}
\DeclareGraphicsExtensions{.pdf,.eps,.png,.jpg,.mps}

\usepackage{amsmath}
\usepackage{amsfonts}
\usepackage{rotating}
\usepackage{multirow}

\usepackage{pifont}
\newcommand*\rot{\rotatebox{90}}

\begin{document}
\begin{titlepage}
	
\centerline{\Large \bf Credit risk and companies' inter-organizational networks:  }
\centerline{\Large \bf Assessing impact of suppliers and buyers on CDS spreads}
\medskip
		
\centerline{Tore Opsahl\footnote{\, The views expressed in this publication do not necessarily reﬂect the views of the authors' employer; Email: \texttt{tore@opsahl.co.uk}} and William Newton}

\medskip
\centerline{December 31, 2015}			
			
\bigskip
\medskip

\begin{abstract}
Companies do not operate in a vacuum. As companies move towards an increasingly specialized production function and their reach is becoming truly global, their aptitude in managing and shaping their inter-organizational network is a determining factor in measuring their health. Current models of company financial health often lack variables explaining the inter-organizational network, and as such, assume that (1) all networks are the same and (2) the performance of partners do not impact companies. This paper aims to be a first step in the direction of removing these assumptions. Specifically, the impact is illustrated by examining the effects of customer and supplier concentrations and partners' credit risk on credit-default swap (CDS) spreads while controlling for credit risk and size. We rely upon supply-chain data from Bloomberg that provides insight into companies' relationships. The empirical results show that a well diversified customer network lowers CDS spread, while having stable partners with low default probabilities increase spreads. The latter result suggests that successful companies do not focus on building a stable eco-system around themselves, but instead focus on their own profit maximization at the cost of the financial health of their suppliers' and customers'. At a more general level, the results indicate the importance of considering the inter-organizational networks, and highlight the value of including network variables in credit risk models.
\end{abstract}

\bigskip
\medskip

\noindent{}{\bf Keywords:} CDS, Credit Risk, Networks
\end{titlepage}

\newpage

\section{\label{sec:intro}Introduction}

The inter-organizational network surrounding companies provides opportunities and constrains behavior. By creating an extensive network of customers and suppliers (collectively called partners), a company is likely to increase shareholder value by specializing on core products and services. This feature can be traced all the way back, for workers instead of organizations, to Adam Smith and division of labor. At an organizational level, similar effects are likely to impact companies' performance and financial health. For example, the assembly of electronic products is becoming an increasingly international process, where components and technology are sourced from specialist suppliers. This process enables companies to focus on core skills, where return on investment and shareholder value are likely to be optimized.

A host of academic papers have applied a network perspective to identify and quantify the impact of inter-organizational networks on various success factors \cite{cross03}. According to the network perspective, organizations are embedded within networks of interconnected relationships. Different positions in a network are associated with a range of outcomes, such as imitation, adaptation, innovation, firm survival, and performance \cite{brass04}. For example, companies in a position of brokering between others have relatively higher returns \cite{bae04}.

Unlike traditional perspectives that treat organizations as independent observations and focus simply on their attributes, network science incorporates additional structural information. These two perspectives are not mutually exclusive and can be combined to conduct a superior analysis by including variables based on network characteristics alongside attribute-based ones.

\subsection{Concentration}

The first feature of the inter-organizational network that this article attempts to tackle is concentration. For a company, concentrations can exist both on the customer-side and the supplier-side. These two aspects represents two different types of underlying risks. First, customer-concentration is an immediate threat to the revenue stream as a single or a select few customers provide the majority of revenue received.  For example, Table 1 highlights the top customers of TripAdvisor. As Expedia represents more than a quarter of TripAdvisor's revenue, TripAdvisor is vulnerable to two scenarios: (1) Expedia is in a position to squeeze its profit margins and (2) a substantial chunk of revenue would potentially be affected were Expedia to default, downsize, or restructure. 

\begin{table}
	\centering
	\begin{tabular}{lr}
		\hline
		Customers & \% Revenue \\
		\hline
		Expedia Inc & 25.63\% \\
		Priceline Group Inc/The & 19.83\% \\
		Orbitz Worldwide Inc & 5.44\% \\
		CTRIP.COM International Ltd & 4.79\% \\
		Google Inc & 1.53\% \\
		American Airlines Group Inc & 1.42\% \\
		\hline		
	\end{tabular}
	\caption{Customers with a quantified relationship representing more than 1\% of TripAdvisor Inc's annual revenue}
	\label{tab:TRIP}
\end{table}

Second, concentration might also occur among suppliers. In this case, a default event could threaten business continuity and pose as a source of operational risk, which ultimately could be detrimental to the financial health of a company. Conversely to customer concentration that might imply a squeeze on the focal company's profit margins through revenue reduction, supplier concentration could lead to cost increases. 

\subsection{Influence: Partners' default probability}

A key feature of networks is the relatively high number of ties among similar nodes \cite{mcpherson01}. This feature is often referred to as homophily. Two causal mechanisms lead to homophily: selection and influence \cite{aral09}. First, similar nodes tend to form ties together (i.e., selection). Second, dissimilar nodes tend to become more similar over time if they are tied (i.e., influence). The latter feature is of key interest when understanding the consequences of the network as opposed to the mechanisms underpinning the network. An understanding of influence forms the basis for assessing and quantifying contagion, diffusion, and cascades.  

From an inter-organizational network perspective, the financial health of companies' partners could provide additional insight into their own financial health. Specifically, influence can be thought of as directly improving or worsening a company's health (e.g., investment grade companies are more stable and thereby bring about less volatility to their partners). The lowering in volatility is likely to bring about a smaller risk premium, and as such, enable a smoother operation of the overall system. Conversely, the system might not be driven by global optimization, but by local optimization instead. Companies are likely to prioritize their own profit maximization (i.e., local optimization) at the expense of ensuring a stable overall ecosystem (i.e., global optimization). Given the opposing theoretical aspects, it is an empirical question of whether, and the extent to which, interacting with stable partners is positively or negatively related to financial health of companies.

The remainder of this article is organized as follows. First the methodology, including data collection and metric construction, is presented. This is followed by results. Finally, a conclusion and discussion section ends the paper with notes on general applicability, limitations, and avenues of future work.

\section{Methodology}

To model the effect of customer and supplier concentrations and partners' credit risk on the financial health of companies, we apply a regression framework with credit-default swap (CDS) spreads as the dependent variable while controlling for credit risk and market cap. Our observations are all companies with 5-year CDS spreads on Bloomberg on April 29, 2014. This includes 828 companies. We limit our observations by excluding banks as they “lack a hard asset/manufacturing-type of supply chain”\footnote{Advisory note on the Bloomberg SPLC-screen when analyzing financial institutions}. This implies removing 152 financial companies from the sample. As such, the total number of observations is 676. Additionally, we weight the observations by the reciprocal of companies' logged market cap in billions to account for the fact that larger companies are more likely to be included in the sample than smaller companies. This alleviates potential correlation between the dependant variable and inclusion probability \cite{fuller09} and enables a better understanding of the financial health of all companies instead of simply the ones that are more likely to have an observable 5-year CDS spread.

The sub-sections below provide details on the data collection and metric operationalization for quantifying the impact of the inter-organizational networks on CDS spreads. Note that all variables tend to be skewed and, as shown below for the dependent variable, transforming the variables by the natural logarithm of them lessens the skewness. For count variables (e.g., number of suppliers) with a potential zero score, the log is taken of 1 plus the variable. More details and distribution plots for all variables are found in the Appendix.

\subsection{Dependent variable: CDS spreads}

Success or performance of a company can be quantified by a number of metrics. A traditional risk framework view focuses on the financial health of companies by predicting default probabilities. These probabilities are often arrived at by modeling historical default events in a logistic-regression framework or applications of Merton's structural default model. However, few companies, and especially large public companies, default. As such, these frameworks are hard to calibrate and exposed to rare-event bias. 

To overcome potential biases, we choose an outcome variable that is quantified for a large number of companies and also closely related to the financial health of a company \cite{longstaff05}: CDS spread. Specifically, we use 5-year market CDS spread (in basis points; bps) as listed on Bloomberg under the Default Risk Monitor (DRAM)-view.

\begin{figure}[ht]
	\centering
	\includegraphics[width=.9\linewidth]{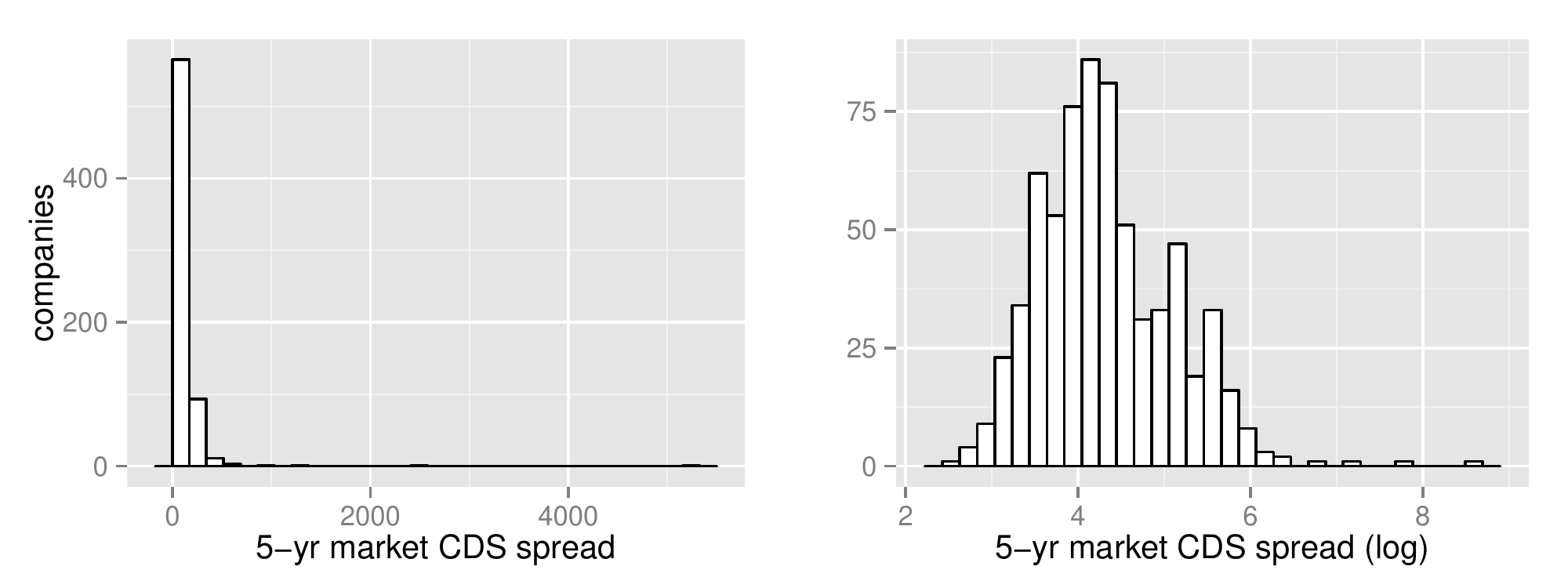}
	\caption{Distribution of 5-year market CDS spreads}
	\label{fig:distCDS}
\end{figure}

The distribution of the 5-year market CDS spreads is highly right-skewed (Figure~\ref{fig:distCDS}a), but conforms closer to a Gaussian shape after transforming it with the natural logarithm (Figure~\ref{fig:distCDS}b). This transformation of the dependent variable enables an ordinary least square (OLS) regression framework to be applied, which lessens the complexity of the model. A well-tested simple model allows us to focus on incorporating novel metrics to assess the impact of the inter-organizational network on the financial health of companies.

\subsection{Inter-organizational network data}

The interdependencies among companies are not straight-forward to measure, assess, and collect. In fact, maintaining the confidentiality of this information might be of strategic importance to the companies. For example, it could enable customers to circumvent the focal company by purchasing directly from its suppliers. Nevertheless, some companies publish their supply chain (e.g., Apple provides a list of their top 200 suppliers\footnote{See apple.com/supplierresponsibility/}) and others' become known through media coverage and quarterly reports. 

One source that attempts to collect the interdependencies among companies is Bloomberg. The supply chain analysis (SPLC)-view gives an insight into the inter-organizational network in which a company operates by listing large suppliers, customers, and peers. The entities in the network are identified by combining Bloomberg analysts' assessment, company reports, quarterly filings, company news releases, and media coverage. To gauge the converage of this data, we performed an ad-hoc analysis of Apple's top 200 suppliers-list. There are 695 suppliers listed in the Bloomberg data, and the intersection among these lists are 191 suppliers, which indicate a 95.5\% coverage rate.

A number of identified relationships are quantified (e.g., Toyota Motors is 3M biggest customer responsible for 4.4\% of their revenue). This information is, however, only populated for about 37\% of the relationships identified\footnote{In total there are 100,030 relationships (63,001 supplier relationships and 37,029 customer relationships) for the included companies in this study. Out of these, 37,014 relationships are quantified (23,497 supplier relationships and 13,517 customer relationships).}. Moreover, the quality of these estimates is uncertain due to this information not being required in regulatory filings.

A key limitation of using this data for network analysis is the difficulty of extracting it for multiple companies. To overcome this limitation, we created a custom tools to aid this otherwise manual process.  

\subsection{Independent variables: Concentration}
\label{sec:varsConcentration}

We consider three features when assessing concentration of customers and suppliers. First, the numbers of customers and suppliers are key to understand how concentrated an inter-organizational network is. This is akin to degree within the network science literature \cite{opsahl10}. An increase in these numbers is likely to suggest additional concentration as only large partners representing concentration are listed in Bloomberg. 

Second, the distribution of relationship values tends to be skewed (e.g., see table 1 for TripAdvisor Inc's customers). Skewness brings about greater concentration that simply the number of partners. 

Third, concentration can also arise through other companies. For example, both TripAdvisor and Expedia are suppliers to American Airlines, and since TripAdvisor and Expedia are mutually connected, American Airlines' are further concentrated than what simply the number of suppliers and skewness would suggest.

\begin{figure}[ht]
	\centering
	\includegraphics[width=.4\linewidth]{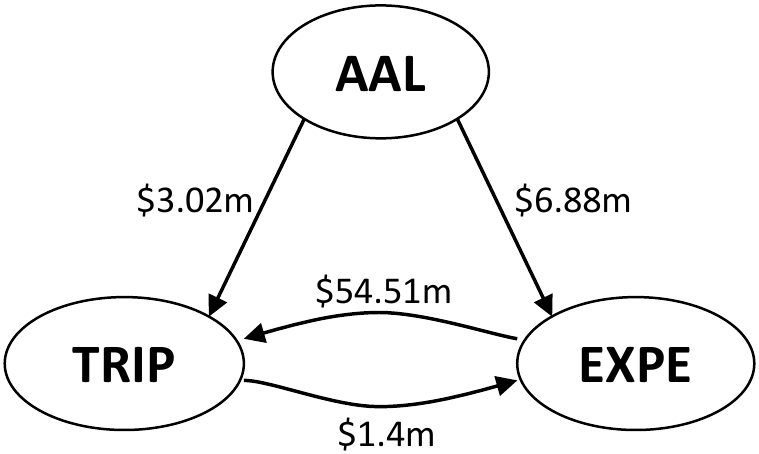}
	\caption{Triad among TripAdvisor Inc, Expedia Inc, and American Airlines Group Inc}
	\label{fig:networkTRIP}
\end{figure}

To measure the concentrations of customers and suppliers, it is common to apply the Herfindahl index. For example, the SEC applies it to assess the competitiveness of sectors and creation of monopolies when considering whether or not to approve mergers. It is defined as the sum of squared proportions. The square ensures that a single large concentration weights more than many smaller ones. In our context, we use the percentage of revenue that a relationship represents for customer-concentration (percentage of total costs for supplier-concentration). In the example of customer-concentration for TripAdvisor (Table 1), the metric is the sum of $0.2563^2$, $0.1983^2$, $0.0544^2$, $0.0479^2$, $0.0153^2$, $0.01422^2$, and so on for all customers, which is about 11\%. Formally, it is:

\begin{equation}
HHI =\sum_j \left( \frac{w_{ij}}{w_{i++}}  \right) ^2 = \sum_j {p_{ij}}^2
\end{equation}
where $w_{ij}$ is the relationship value from company $i$ to company $j$, $w_{i++}$ is the total revenue (cost) for customer (supplier)-analysis, and $p_{ij}$ is the fraction between these two values. 

It is worth noting that the Herfindahl index does not consider the third feature listed above: indirect concentrations. To incorporate this feature, constraint, an advanced version of the Herfindahl index, is often used in network analysis \cite{burt92}. This metric comes from the structural holes literature and is formally defined as:

\begin{equation}
constraint (i) = \sum_j \left( p_{ij} + \sum_q p_{iq} p_{qj} \right) ^2
\end{equation}
where $q$ are companies indirectly connected to company $i$ and company $j$. If there were no indirect connections, this metric would be equal to the Herfindahl index as the $p_{iq}p_{qj}$-term would be equal to 0. 

However, this metric requires a exponentially larger data collection effort as the network of all partner companies would need to be acquired. As such, we have computed five metrics for both customer and supplier lists:

\begin{enumerate}
	\item Number of companies
	\item Number of companies with a Bloomberg identifier, which are likely to be public companies
	\item Number of customers (suppliers) with percentage of revenue (cost) defined
	\item The sum of percentage of revenue (cost) that customers (suppliers) represent
	\item The Herfindahl Index
\end{enumerate}

It is worth noting that the last two are only calculated on the subset of partners with a quantified relationship. As such, these metrics only consider about 37\% of the relational data.

\subsection{Independent variables: Influence}

We consider the default probability of partners to test whether the financial health of partners impact the focal company. Specifically, we take the average of customers' (suppliers') Bloomberg-defined default probability derived from a structural Merton model. We chose this variable instead of CDS spreads as CDS spreads are only available for a limited sample of companies, and as such, would lead to sparse data issues and dropped observations. On the contrary, a default probability is available for the partners of 70\% of relationships\footnote{Out of the total number of relationships, the partners of 69,637 relationships have a default probability (45,844 supplier relationships and 23,793 customer relationships).}. 

For both suppliers and customers, we take a simple average and a weighted one. The weights are based on the percentage of revenue (costs) that the particular customer (supplier) represents. While the simple average is applied to about 70\% of relationships, the weighted is only calculated on 21\% (12\%) of customer (supplier) relationships as the percentage of revenue (cost) is only available for 37\% of relationships. 

\subsection{Control variables}

To control for general financial information, we include the corporate default probability as derived by Bloomberg's structural Merton model \cite{bbgDRSK}. This variable is correlated with CDS spreads (pair-wise correlation of 0.40; $R^2$ is 0.16 in a univariate model). In fact, it can be argued that these two variables are the same. However, the model-derived default probability only considers the independent variables used in the model, and the purpose of this paper is to highlight that inter-organizational network variables have the potential of increasing the explanatory power in a combined model.

Additionally, we include the market cap of companies as a proxy of size and the liquidity of the swaps. Larger companies are likely to provide more information to the market, be rated by a larger set of investors, and have more contracts with various maturity dates than smaller ones. As such, their CDS contracts are likely to be traded more frequently, which increases the liquidity.

We further control for the Global Industry Classification Standard (GICS) sector of the companies. This ensures that sectoral differences are parsed out. 

Finally, indicator variables for the country of risk associated with the companies are included. This variable attempts to overcome spread differences due to countries. The country of risk is determined based on ``four factors listed in order of importance: management location, country of primary listing, country of revenue and reporting currency of the issuer. Management location is defined by country of domicile unless location of such key players as Chief Executive Officer (CEO), Chief Financial Officer (CFO), Chief Operating Officer (COO), and/or General Counsel is proven to be otherwise''. 

\section{Results}

The regression results from a select set of combination of variables are listed in Table~\ref{tab:results}. Model 1 is a baseline model without any inter-organizational variables. We find a strong link between default probability and CDS spreads. Together with country and sector indicator variables and market cap, this baseline or control model explains 68\% of the variance in CDS spreads. In the Appendix, descriptive statistics and pair-wise correlation for the main variables are listed in Table~\ref{tab:pwcorr} and results for the indicator variables are in Table~\ref{tab:resultsFull} .

\begin{table}[ht]
	\centering
	\begin{tabular}{lr@{}lr@{}lr@{}l}
		\hline
		& \multicolumn{6}{c}{Models} \\
		\cline{2-7}
		Variables & \multicolumn{2}{c}{M1} & \multicolumn{2}{c}{M2} & \multicolumn{2}{c}{M3} \\
		\hline
		\textit{Concentration} & & & & & & \\
		~~Suppliers (log) & & & -0.081 & ** & -0.042 & \\
		& & & (0.027 & ) & (0.026 & ) \\
		~~Customers (log) & & & 0.135 & *** & 0.114 & *** \\
		& & & (0.022 & ) & (0.022 & ) \\
		\textit{Influence} & & & & & & \\ 
		~~Suppliers (log) & & & & & -0.251 & *** \\
		& & & & & (0.033 & ) \\
		~~Customers (log) & & & & & -0.077 & ** \\
		& & & & & (0.029 & ) \\
		DP (log) & 0.315 & *** & 0.363 & *** & 0.396 & *** \\
		& (0.022 & ) & (0.024 & ) & (0.023 & ) \\
		Market Cap (log) & -0.007 &  & -0.026 &  & -0.020 &  \\
		& (0.019 & ) & (0.025 & ) & (0.023 & ) \\
		\textit{GICS sector indicators} &  \multicolumn{2}{c}{\textit{incl.}} & \multicolumn{2}{c}{\textit{incl.}} & \multicolumn{2}{c}{\textit{incl.}}  \\
		\textit{Country of Risk indicators} & \multicolumn{2}{c}{\textit{incl.}} & \multicolumn{2}{c}{\textit{incl.}} & \multicolumn{2}{c}{\textit{incl.}}  \\
		Constant & 7.284 & *** & 7.538 & *** & 5.599 & *** \\
		& (0.159 & ) & (0.201 & ) & (0.303 & ) \\
		\hline
		Observations & 676 &  & 676 &  & 676 & \\
		$R^2$ & 0.6806 &  & 0.6997 &  & 0.7319 & \\
		Adjusted $R^2$ & 0.6642 &  & 0.6832 &  & 0.7163 & \\
		$\Delta R^2$ (bps; from M1)  &  &  & 191 &  & 513 & \\
		\hline
	\end{tabular}
	\caption{Regression results; Full table available in the Appendix. \newline $^*p<0.05; ^{**}p<0.01; ^{***}p<0.001$}
	\label{tab:results}
\end{table}

We find a relationship between CDS spreads and the various inter-organizational variables. Specifically, Model 2 shows that having many large suppliers lowers the CDS spread while having large customers increase the spread. The latter feature is maintained in Model 3 when including influence variables. Both influence variables are negatively related to spread. This indicates that higher default probabilities of partners are associated with lower spreads of the focal company. This effect suggests that financially healthy focal companies prioritize profit maximization over overall stability in their inter-organizational network by, for example, squeezing their suppliers' profit margins, which in turn is detrimental to their financial health. For example, Walmart has the potential exerting pressure on suppliers if it represents a large proportion of their revenue. 
As such, it does seem that local optimization is favored over global optimization.

The inter-organizational variables increase the explanatory power of the framework. By including concentration and influence variables, the $R^2$ increases from 0.68 to 0.73. This 513bps increase suggests that inter-organizational variables provide novel and additional insight into CDS spreads.

For a more comprehensive set of variable-combinations, see Table~\ref{tab:resultsIntermediate} in the Appendix. It is worth noting that some of these combinations bring about greater model improvement, but we have chosen the simpler operationalizations of concentration and influence in Table~\ref{tab:results}. To ensure an identical sample, we set the suppliers' (customers') average default probability to the average of observed values when missing as 61 companies have either no suppliers or no customers with a defined default probability. An alternative to Model 3 with these observations dropped increases the  $R^2$ to 0.8220. To be conservative, we choose the identical larger sample used in Models 1 and 2 with lower model improvement for comparability. As a robustness check, the analysis was conducted on the complete set of observation (i.e., including financial institutions) and the model improvement was maintained albeit with smaller  $R^2$ values (m1: 0.6231; m2: 0.6300; m3: 0.6585; $\Delta$m2: 69bps; $\Delta$m3: 354bps). 

\section{Conclusion and discussion}

This project has shown that the inclusion of inter-organizational network variables increases the explanatory power of models predicting the CDS spreads, and in general the financial health of companies. We applied a simple OLS regression framework and found an increase in $R^2$ of 191bps and 513bps when including concentration and influence variables, cumulatively. These results have direct applicability to credit risk framework, and suggests that they can be improved by including inter-organizational network information. 

The analysis performed in this paper has a number of limitations. Chief among those is the simplicity of the analysis performed. It would surely be improved in more advanced models. For example, we applied a static framework, but spreads vary over time and the volatility could be modeled. Additionally, the supply-chain data used is solely available for public companies. While attempting to mitigate inclusion probability, this limits the applicability of the analysis. Moreover, we did not collect sufficient data to analyze network constraint due to the non-incremental effect of the Herfindahl index over the number of relationships. Finally, although CDS spreads are market variables, they are impacted by algorithmic trading. In turn, this might imply that they are converging around an aggregation of the various trading algorithms used.

A number of avenues of future work exist. We are particularly interested in appending the inter-organizational network variables to other existing credit risk frameworks, such as probability of default models. This would enable an understanding of whether network variables do increase the explanatory power in advanced frameworks predicting the default likelihood.

\clearpage
\appendix

\section*{Appendices}

\section{Variable distributions}

\begin{figure}[!ht]
	\centering
	\includegraphics[width=.7\linewidth]{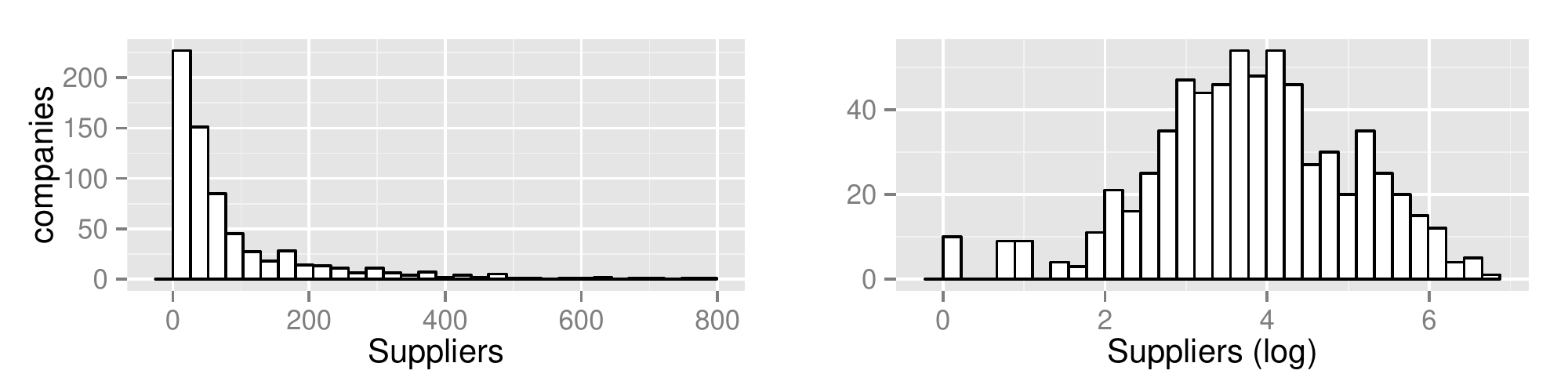}
	\includegraphics[width=.7\linewidth]{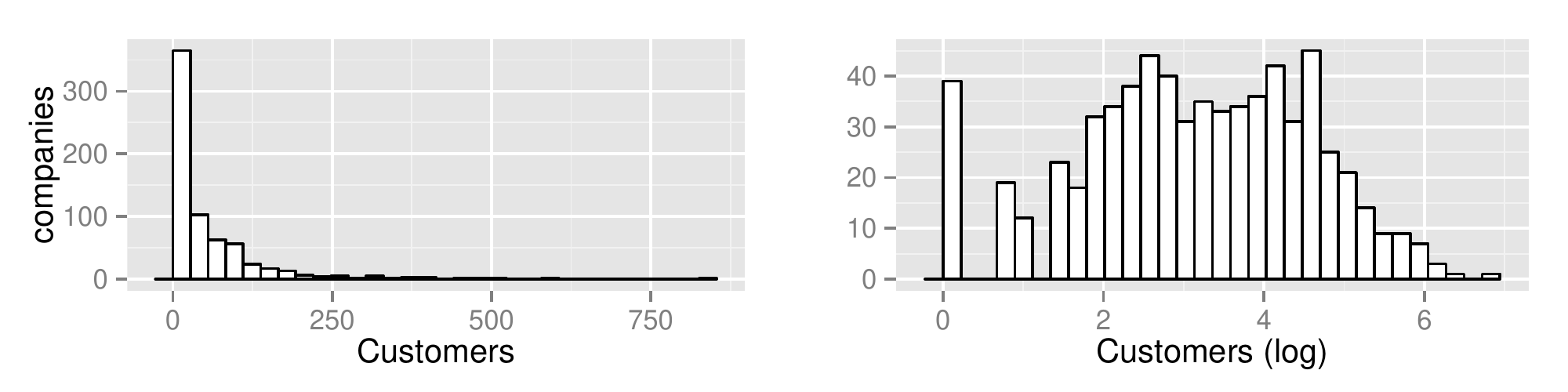}
	\includegraphics[width=.7\linewidth]{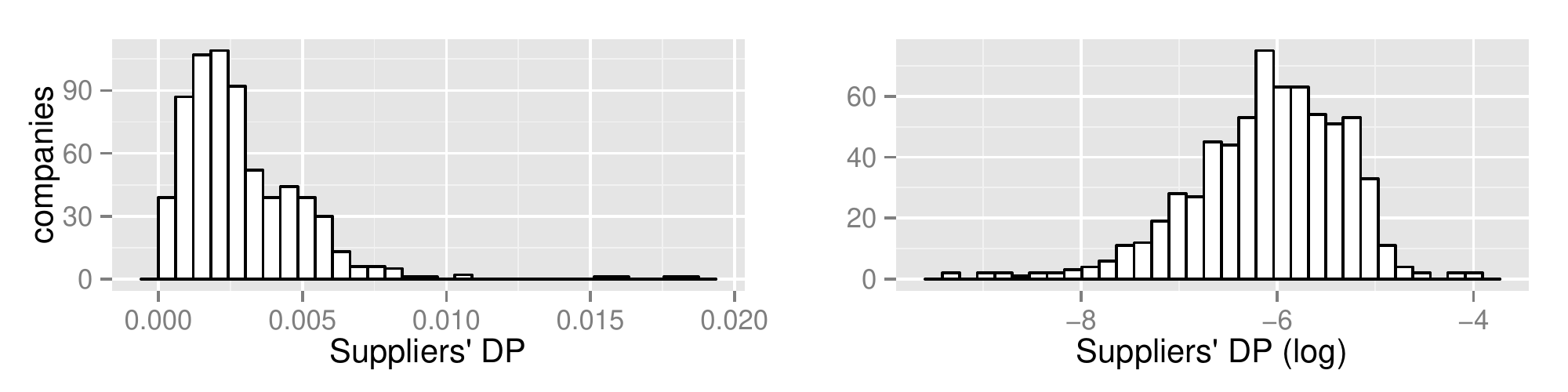}
	\includegraphics[width=.7\linewidth]{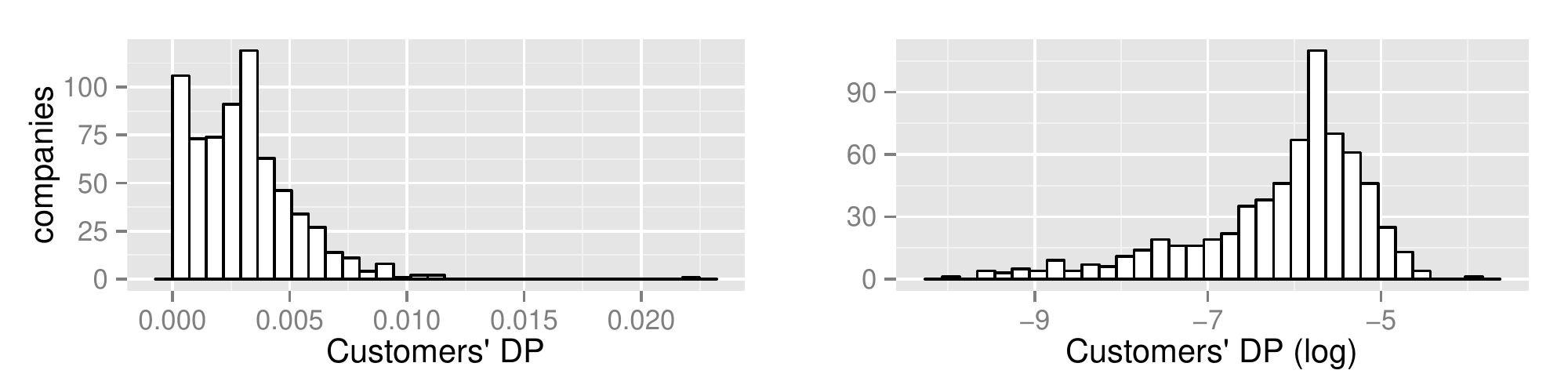}
	\includegraphics[width=.7\linewidth]{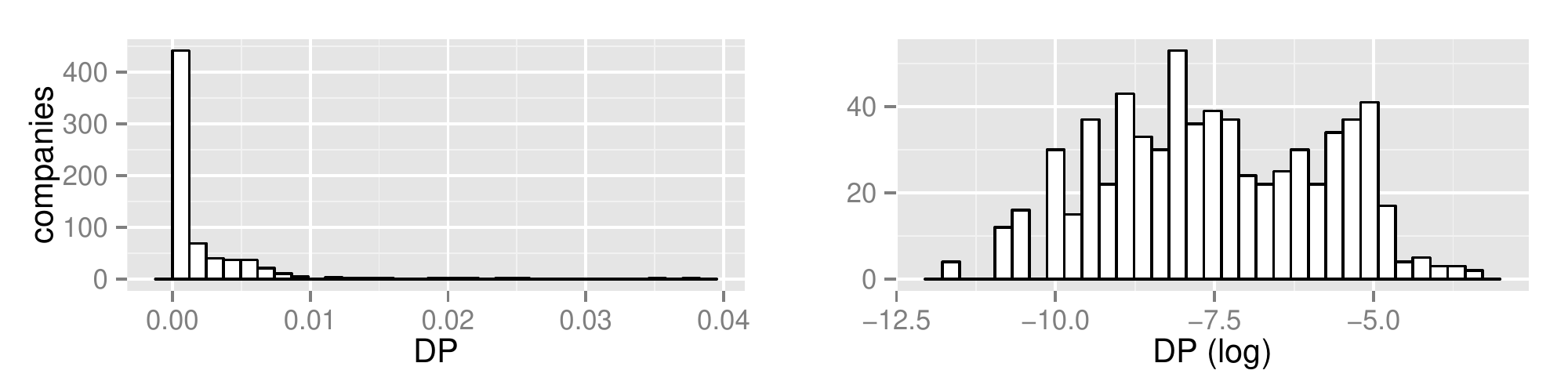}
	\includegraphics[width=.7\linewidth]{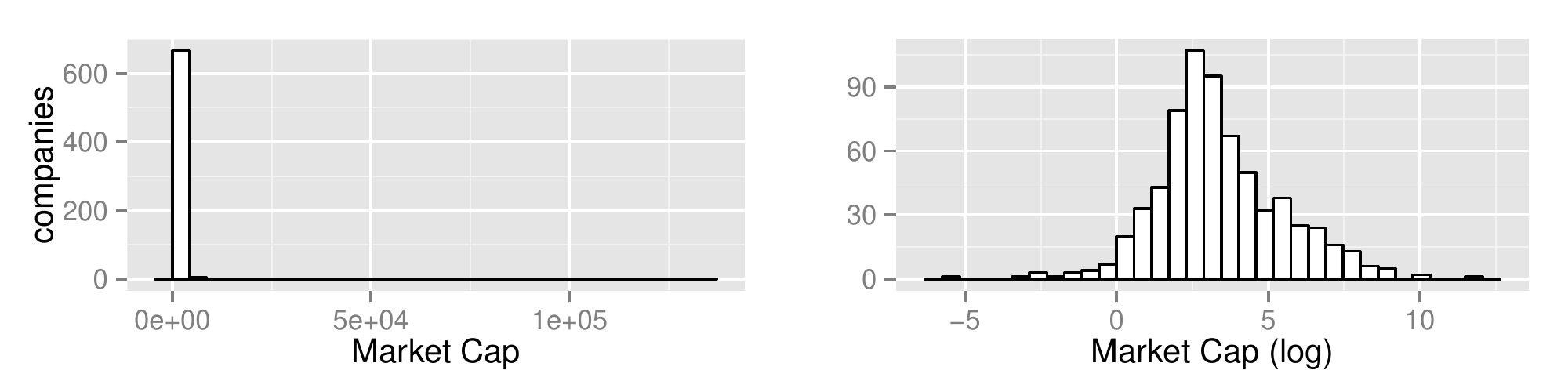}
	\caption{Distributions of main independent variables untransformed and when transformed with the natural logarithm}
	\label{fig:distVars}
\end{figure}

Figure~\ref{fig:distCDS} showed that the distribution of the dependent variable is highly skewed and become more Gaussian-like if the variable is transformed with the natural logarithm. Figure~\ref{fig:distVars} shows the effect of similar transforms for the main independent variables.

\clearpage

\section{Variable statistics}
\begin{table}[!ht]
	\centering
		\rot{
	\begin{tabular}{l|rr|rrrrrrr}
		& & & \multicolumn{6}{c}{Pair-wise correlations} \\
		\cline{4-10}
		Variable & \multicolumn{1}{|c}{Mean} & \multicolumn{1}{c|}{SD} & \multicolumn{1}{|c}{1}  & \multicolumn{1}{c}{2}  & \multicolumn{1}{c}{3}  & \multicolumn{1}{c}{4}  & \multicolumn{1}{c}{5}  & \multicolumn{1}{c}{6}  & \multicolumn{1}{c}{7}    \\
		\hline
		1. CDSmkt (log) & 4.33 & 0.78 & 1.00 &  &  &  &  &  &  \\
		2. Suppliers (log) & 3.78 & 1.27 & -0.33 & 1.00 &  &  &  & &   \\
		3. Customers (log) & 3.12 & 1.46 & -0.34 & \textbf{0.60} & 1.00 &  &  &  &  \\
		4. Suppliers' avg. DP (log) & -6.10 & 0.78 & -0.18 & 0.18 & 0.27 & 1.00 &  &  &  \\
		5. Customers' avg. DP (log) & -6.17 & 1.02 & -0.11 & 0.20 & 0.23 & \textbf{0.44} & 1.00 &  &  \\
		6. DP (log) & -7.50 & 1.72 & \textbf{0.40} & -0.11 & -0.07 & 0.38 & 0.35 & 1.00 &  \\
		7. Market Cap (log) & 3.39 & 2.08 & \textbf{-0.51} & \textbf{0.57} & \textbf{0.49} & \textbf{0.41} & 0.35 & 0.15 & 1.00  \\
		\hline	
	\end{tabular}}
	\caption{Means, standard deviations, and pair-wise correlations for the main independent variables. Absolute inter-variable correlations above 0.4 are highlighted in bold.}
	\label{tab:pwcorr}
\end{table}

\clearpage

\section{Full regression table}

\begin{table}[!ht]
	\centering
	\begin{footnotesize}
	\begin{tabular}{llr@{}llr@{}llr@{}ll}
		\hline
		\multicolumn{2}{l}{Variables} & \multicolumn{3}{c}{Model 1} & \multicolumn{3}{c}{Model 2} & \multicolumn{3}{c}{Model 3} \\
		\hline
		\multicolumn{2}{l}{\textit{Concentration}} & & & & & & & & & \\
		& Suppliers (log) & & & & -0.081 & ** & (0.027) & -0.042 & & (0.026) \\
		& Customers (log) & & & & 0.135 & *** & (0.022) & 0.114 & *** & (0.022) \\
		\hline
		\multicolumn{2}{l}{\textit{Influence}} & & & & & & & & & \\ 
		& Suppliers (log) & & & & & & & -0.251 & *** & (0.033) \\
		& Customers (log) & & & &  & & & -0.077 & ** & (0.029) \\
		\hline
		\multirow{8}{*}{\rot{GICS sectors}} & Consumer Discretionary & 0.408 & *** & (0.089) & 0.494 & *** & (0.087) & 0.463 & *** & (0.083) \\
		& Consumer Staples & -0.052 &  & (0.120) & 0.088 &  & (0.118) & -0.056 &  & (0.113) \\
		& Energy & 0.126 &  & (0.116) & 0.184 &  & (0.115) & 0.193 &  & (0.110) \\
		& Health Care & -0.152 &  & (0.158) & -0.031 &  & (0.155) & -0.137 &  & (0.147) \\
		& Information Technology & -0.826 & *** & (0.114) & -0.940 & *** & (0.114) & -0.762 & *** & (0.110) \\
		& Materials & 0.062 &  & (0.106) & 0.060 &  & (0.103) & 0.018 &  & (0.098) \\
		& Telecommunication Srv & -0.007 &  & (0.166) & 0.076 &  & (0.162) & -0.050 &  & (0.154) \\
		& Utilities & -0.075 &  & (0.131) & 0.055 &  & (0.130) & 0.006 &  & (0.124) \\
		\hline
		\multirow{23}{*}{\rot{Country of Risk indicators (ref: US)}} & Austria & -0.573 &  & (0.584) & -0.379 &  & (0.568) & -0.558 &  & (0.542) \\
		& Belgium & -0.675 &  & (0.444) & -0.568 &  & (0.432) & -0.457 &  & (0.409) \\
		& Brazil & -0.348 &  & (0.770) & -0.386 &  & (0.748) & -0.205 &  & (0.709) \\
		& Canada & -0.116 &  & (0.184) & -0.010 &  & (0.180) & 0.092 &  & (0.171) \\
		& Denmark & -0.498 &  & (0.679) & -0.452 &  & (0.661) & -0.492 &  & (0.625) \\
		& Finland & 0.190 &  & (0.254) & 0.256 &  & (0.247) & 0.335 &  & (0.235) \\
		& France & -0.478 & *** & (0.097) & -0.462 & *** & (0.099) & -0.569 & *** & (0.099) \\
		& Germany & -1.313 & *** & (0.097) & -1.331 & *** & (0.094) & -0.974 & *** & (0.098) \\
		& Greece & -0.301 & ** & (0.115) & -0.399 & *** & (0.116) & -0.178 &  & (0.114) \\
		& Indonesia & -0.326 &  & (1.117) & -0.409 &  & (1.093) & -0.003 &  & (1.036) \\
		& Ireland & 0.013 &  & (0.793) & 0.189 &  & (0.772) & -0.004 &  & (0.731) \\
		& Italy & -0.352 & * & (0.159) & -0.593 & *** & (0.160) & -0.533 & *** & (0.151) \\
		& Japan & -1.321 & *** & (0.127) & -1.388 & *** & (0.134) & -1.130 & *** & (0.131) \\
		& Luxembourg & -0.277 &  & (0.553) & -0.401 &  & (0.538) & -0.345 &  & (0.509) \\
		& Mexico & -0.351 &  & (0.667) & -0.423 &  & (0.651) & -0.285 &  & (0.617) \\
		& Netherlands & -0.616 & ** & (0.233) & -0.568 & * & (0.227) & -0.478 & * & (0.215) \\
		& Norway & 1.595 & *** & (0.348) & 1.608 & *** & (0.339) & 1.882 & *** & (0.322) \\
		& Peru & -0.606 &  & (0.731) & -0.680 &  & (0.712) & -0.576 &  & (0.674) \\
		& Portugal & 0.228 &  & (0.331) & 0.216 &  & (0.322) & 0.309 &  & (0.305) \\
		& Spain & 0.029 &  & (0.253) & 0.013 &  & (0.247) & -0.097 &  & (0.235) \\
		& Sweden & -0.527 &  & (0.314) & -0.480 &  & (0.306) & -0.452 &  & (0.290) \\
		& Switzerland & -0.339 &  & (0.269) & -0.319 &  & (0.261) & -0.159 &  & (0.248) \\
		& United Kingdom & -0.257 & * & (0.117) & -0.199 &  & (0.114) & -0.186 & & (0.108) \\
		\hline
		\multicolumn{2}{l}{DP} & 0.315 & *** & (0.022) & 0.363 & *** & (0.024) & 0.396 & *** & (0.023) \\
		\multicolumn{2}{l}{Market Cap} & -0.007 &  & (0.019) & -0.026 &  & (0.025) & -0.020 &  & (0.023) \\
		\multicolumn{2}{l}{Constant} & 7.284 & *** & (0.159) & 7.538 & *** & (0.201) & 5.599 & *** & (0.303) \\
		\hline
		\multicolumn{2}{l}{Observations} & \multicolumn{2}{r}{676} & & \multicolumn{2}{r}{676} & &\multicolumn{2}{r}{676} & \\
		\multicolumn{2}{l}{$R^2$} & \multicolumn{2}{r}{0.680} & & \multicolumn{2}{r}{0.6997} & & \multicolumn{2}{r}{0.7319} & \\
		\multicolumn{2}{l}{Adjusted $R^2$} & \multicolumn{2}{r}{0.6642} & & \multicolumn{2}{r}{0.6832} & & \multicolumn{2}{r}{0.7163} & \\
		\multicolumn{2}{l}{$\Delta R^2$ (bps; from M1)}  &  \multicolumn{2}{r}{N/A} & & \multicolumn{2}{r}{191} & & \multicolumn{2}{r}{513} & \\
		\hline
	\end{tabular}
	\caption{Regression table with all indicator control variables. The reference GICS sector is industrials. $^*p<0.05; ^{**}p<0.01; ^{***}p<0.001$}
	\label{tab:resultsFull}
\end{footnotesize}
\end{table}

\clearpage

\section{Intermediate regression tables}
\begin{table}[!ht]
	\centering
	\begin{tiny}
		\rot{
		\begin{tabular}{l|l|l|r@{}l|r@{}l@{}r@{}l@{}r@{}l@{}r@{}l@{}r@{}l@{}r@{}l@{}r@{}l@{}r@{}l@{}r@{}l@{}r@{}l}
			\hline
			\multicolumn{3}{c|}{~} & \multicolumn{2}{|c|}{Control} & \multicolumn{20}{|c}{Control model with each of the concentration variables separately} \\
			\hline
			\multicolumn{3}{l|}{Variables} & \multicolumn{2}{c|}{Model 1} & \multicolumn{2}{c}{Model 2} & \multicolumn{2}{c}{Model 3} & \multicolumn{2}{c}{Model 4} & \multicolumn{2}{c}{Model 5} & \multicolumn{2}{c}{Model 6} & \multicolumn{2}{c}{Model 7} & \multicolumn{2}{c}{Model 8} & \multicolumn{2}{c}{Model 9} & \multicolumn{2}{c}{Model 10} & \multicolumn{2}{c}{Model 11} \\
			\hline
			\multirow{20}{*}{\rot{Concentration}} & \multirow{10}{*}{\rot{Suppliers}} & Metric 1 &  &  & -0.049 &  &  &  &  &  &  &  &  &  &  &  &  &  &  &  &  &  &  & \\
			& & & & & (0.027 & ) &  &  &  &  &  &  &  &  &  &  &  &  &  &  &  &  &  & \\
			& & Metric 2 &  &  &  &  & -0.068 & * &  &  &  &  &  &  &  &  &  &  &  &  &  &  &  & \\
			& & & & & & & (0.028 & ) &  &  &  &  &  &  &  &  &  &  &  &  &  &  &  & \\
			& & Metric 3 &  &  &  &  &  &  & -0.024 &  &  &  &  &  &  &  &  &  &  &  &  &  &  & \\
			& & & & & & & & & (0.027 & ) &  &  &  &  &  &  &  &  &  &  &  &  &  & \\
			& & Metric 4 &  &  &  &  &  &  &  &  & -0.085 & *** &  &  &  &  &  &  &  &  &  &  &  & \\
			& & & & & & & & & & & (0.021 & ) &  &  &  &  &  &  &  &  &  &  &  & \\
			& & Metric 5 &  &  &  &  &  &  &  &  &  &  & 1.004 &  &  &  &  &  &  &  &  &  &  & \\
			& & & & & & & & & & & &  & (0.729 & ) &  &  &  &  &  &  &  &  &  & \\
			\cline{2-25}
			& \multirow{10}{*}{\rot{Customers}} & Metric 1 &  &  &  &  &  &  &  &  &  &  &  &  & 0.122 & *** &  &  &  &  &  &  &  & \\
			& & & & & & & & & & & & & & & (0.022 & ) &  &  &  &  &  &  &  & \\
			& & Metric 2 &  &  &  &  &  &  &  &  &  &  &  &  &  &  & 0.116 & *** &  &  &  &  &  & \\
			& & & & & & & & & & & & & & & &  & (0.023 & ) &  &  &  &  &  & \\
			& & Metric 3 &  &  &  &  &  &  &  &  &  &  &  &  &  &  &  &  & 0.068 & ** &  &  &  & \\
			& & & & & & & & & & & & & & & & & &  & (0.023 & ) &  &  &  & \\
			& & Metric 4 &  &  &  &  &  &  &  &  &  &  &  &  &  &  &  &  &  &  & 0.075 & *** &  & \\
			& & & & & & & & & & & & & & & & & & & & & (0.021 & ) &  & \\
			& & Metric 5 &  &  &  &  &  &  &  &  &  &  &  &  &  &  &  &  &  &  &  &  & 0.026 & \\
			& & & & & & & & & & & & & & & & & & & & & &  & (0.491 & ) \\
			\hline
			\multirow{8}{*}{\rot{Influence}} & \multirow{4}{*}{\rot{Supp.}} & Average & & & & & & & & & & & & & & & & & & \\
			& & & & & & & & & & & & & & & & & & & & & & & \\
			& & W. avg. &  &  &  &  &  &  &  &  &  &  &  &  &  &  &  &  &  &  &  &  &  & \\
			& & & & & & & & & & & & & & & & & & & & & & & \\
			\cline{2-25}
			& \multirow{4}{*}{\rot{Cust.}} & Average &  &  &  &  &  &  &  &  &  &  &  &  &  &  &  &  &  &  &  &  &  & \\
			& & & & & & & & & & & & & & & & & & & & & & & \\
			& & W. avg. &  &  &  &  &  &  &  &  &  &  &  &  &  &  &  &  &  &  &  &  &  & \\
			& & & & & & & & & & & & & & & & & & & & & & & \\
			\hline
			\multicolumn{3}{l|}{DP} & 0.315 & *** & 0.332 & *** & 0.341 & *** & 0.324 & *** & 0.349 & *** & 0.313 & *** & 0.333 & *** & 0.327 & *** & 0.317 & *** & 0.328 & *** & 0.315 & *** \\
			\multicolumn{3}{l|}{~} & (0.022 & ) & (0.024 & ) & (0.024 & ) & (0.024 & ) & (0.023 & ) & (0.022 & ) & (0.022 & ) & (0.022 & ) & (0.022 & ) & (0.022 & ) & (0.022 & )\\
			\multicolumn{3}{l|}{Market Cap} & -0.007 &  & 0.021 &  & 0.034 &  & 0.007 &  & 0.017 &  & -0.008 &  & -0.066 & ** & -0.065 & ** & -0.033 &  & -0.019 &  & -0.007 & \\
			\multicolumn{3}{l|}{~}  & (0.019 & ) & (0.024 & ) & (0.025 & ) & (0.024 & ) & (0.020 & ) & (0.019 & ) & (0.021 & ) & (0.022 & ) & (0.021 & ) & (0.019 & ) & (0.019 & ) \\
			\multicolumn{3}{l|}{GICS sector indi.} & Incl. &  & Incl. &  & Incl. &  & Incl. &  & Incl. &  & Incl. &  & Incl. &  & Incl. &  & Incl. &  & Incl. &  & Incl. & \\
			\multicolumn{3}{l|}{Country of Risk indi.} & Incl. &  & Incl. &  & Incl. &  & Incl. &  & Incl. &  & Incl. &  & Incl. &  & Incl. &  & Incl. &  & Incl. &  & Incl. & \\
			\multicolumn{3}{l|}{Constant} & 7.284 & *** & 7.526 & *** & 7.631 & *** & 7.386 & *** & 7.668 & *** & 7.249 & *** & 7.156 & *** & 7.158 & *** & 7.22 & *** & 7.258 & *** & 7.285 & *** \\
			\multicolumn{3}{l|}{~} & (0.159 & ) & (0.206 & ) & (0.212 & ) & (0.195 & ) & (0.185 & ) & (0.161 & ) & (0.157 & ) & (0.158 & ) & (0.160 & ) & (0.158 & ) & (0.159 & ) \\
			\hline
			\multicolumn{3}{l|}{Observations} & 676 &  & 676 &  & 676 &  & 676 &  & 676 &  & 676 &  & 676 &  & 676 &  & 676 &  & 676 &  & 676 & \\
			\multicolumn{3}{l|}{Dropped obs. (M1)} & N/A &  & 0 &  & 0 &  & 0 &  & 0 &  & 0 &  & 0 &  & 0 &  & 0 &  & 0 &  & 0 & \\
			\multicolumn{3}{l|}{$R^2$} & 0.6806 &  & 0.6823 &  & 0.6836 &  & 0.681 &  & 0.6882 &  & 0.6815 &  & 0.6954 &  & 0.6929 &  & 0.6848 &  & 0.6870 &  & 0.6806 & \\
			\multicolumn{3}{l|}{Adjusted R2} & 0.6642 &  & 0.6654 &  & 0.6668 &  & 0.6641 &  & 0.6716 &  & 0.6647 &  & 0.6792 &  & 0.6766 &  & 0.6680 &  & 0.6704 &  & 0.6637 & \\
			\multicolumn{3}{l|}{$\Delta R^2$ (bps; M1)} & N/A &  & 17 &  & 30 &  & 4 &  & 76 &  & 9 &  & 148 &  & 123 &  & 42 &  & 64 &  & 0 & \\
			\hline
		\end{tabular}}
		\caption{Metrics refer to the list in Section~\ref{sec:varsConcentration}. Standard deviations in parentheses. $^*p<0.05; ^{**}p<0.01; ^{***}p<0.001$}
		\label{tab:resultsIntermediate}
	\end{tiny}
\end{table}

\clearpage

\begin{table}[!ht]
	\centering
	\begin{tiny}
		\rot{
			\begin{tabular}{l|l|l|r@{}l|r@{}l@{}r@{}l@{}r@{}l|r@{}l@{}r@{}l@{}r@{}l@{}r@{}l|r@{}l@{}r@{}l}
				\hline
				\multicolumn{3}{c|}{~} & \multicolumn{2}{|c|}{Control} & \multicolumn{6}{|c}{Concentration model}  & \multicolumn{8}{|c}{Concentration model and single influence variables} & \multicolumn{4}{|c}{Full model}\\
				\hline
				\multicolumn{3}{l|}{Variables} & \multicolumn{2}{c|}{Model 1} & \multicolumn{2}{c}{Model 12} & \multicolumn{2}{c}{Model 13} & \multicolumn{2}{c|}{Model 14} & \multicolumn{2}{c}{Model 15} & \multicolumn{2}{c}{Model 16} & \multicolumn{2}{c}{Model 17} & \multicolumn{2}{c}{Model 18} & \multicolumn{2}{|c}{Model 19} & \multicolumn{2}{c}{Model 20} \\
				\hline
				\multirow{12}{*}{\rot{Concentration}} & \multirow{6}{*}{\rot{Suppliers}} & Metric 1 &  &  & -0.081 & ** &  &  &  &  & -0.051 & * & 0.141 & *** & -0.064 & * & 0.182 & *** & -0.042 &  & 0.116 & ***  \\
				& & & & & (0.027 & ) &  &  &  &  & (0.026 & ) & (0.034 & ) & (0.027 & ) & (0.031 & ) & (0.026 & ) & (0.034 & ) \\ 
				& & Metric 2 &   &  &  &  & -0.103 & *** &  &  &  &  &  &  &  &  &  &  &  &  &  &   \\
				& & &  &  &  &  & (0.028 & ) &  &  &  &  &  &  &  &  &  &  &  &  &  &  \\
				& & Metric 3 &   &  &  &  &  &  & -0.046 &  &  &  &  &  &  &  &  &  &  &  &  & \\
				& & &  &  &  &  &  &  & (0.027 & ) &  &  &  &  &  &  &  &  &  &  &  & \\
				\cline{2-23}
				& \multirow{6}{*}{\rot{Customers}} & Metric 1 &   &  & 0.135 & *** &  &  &  &  & 0.128 & *** & 0.110 & *** & 0.112 & *** & -0.036 &  & 0.114 & *** & 0.010 &   \\
				& & & &  & (0.022 & ) &  &  &  &  & (0.021 & ) & (0.024 & ) & (0.023 & ) & (0.030 & ) & (0.022 & ) & (0.031 & )  \\
				& & Metric 2 &   &  &  &  & 0.135 & *** &  &  &  &  &  &  &  &  &  &  &  &  &  \\
				& & &   &  &  &  & (0.023 & ) &  &  &  &  &  &  &  &  &  &  &  &  &  & \\
				& & Metric 3 &   &  &  &  &  &  & 0.077 & ** &  &  &  &  &  &  &  &  &  &  &  &   \\
				& & &  &  &  &  &  &  & (0.024 & ) &  &  &  &  &  &  &  &  &  &  &  &   \\
				\hline
				\multirow{8}{*}{\rot{Influence}} & \multirow{4}{*}{\rot{Supp.}} & Average &  &  &  &  &  &  &  &  & -0.268 & *** &  &  &  &  &  &  & -0.251 & *** &  & \\
				& & &  &  &  &  &  &  &  &  & (0.032 & ) &  &  &  &  &  &  & (0.033 & ) &  &   \\
				& & W. avg. &  &  &  &  &  &  &  &  &  &  & -0.110 & *** &  &  &  &  &  &  & 0.000 &   \\
				& & &  &  &  &  &  &  &  &  &  &  & (0.025 & ) &  &  &  &  &  &  & (0.025 & ) \\
				\cline{2-23}
				& \multirow{4}{*}{\rot{Cust.}} & Average &  &  &  &  &  &  &  &  &  &  &  &  & -0.119 & *** &  &  & -0.077 & ** &  &    \\
				& & &  &  &  &  &  &  &  &  &  &  &  &  & (0.029 & ) &  &  & (0.029 & ) &  & \\
				& & W. avg. &  &  &  &  &  &  &  &  &  &  &  &  & & & -0.045 & * & & & -0.048 & * \\
				&  &  &  & & & &  &  &  &  &  &  &  &  &  &  & (0.023 & ) & & & (0.023 & ) \\
				\hline
				\multicolumn{3}{l|}{DP} & 0.315 & *** & 0.363 & *** & 0.369 & *** & 0.334 & *** & 0.395 & *** & 0.457 & *** & 0.367 & *** & 0.487 & *** & 0.396 & *** & 0.474 & *** \\
				\multicolumn{3}{l|}{~} & (0.022 & ) & (0.024 & ) & (0.024 & ) & (0.024 & ) & (0.023 & ) & (0.023 & ) & (0.023 & ) & (0.022 & ) & (0.023 & ) & (0.022 & ) \\
				\multicolumn{3}{l|}{Market Cap} & -0.007 &  & -0.026 &  & -0.012 &  & -0.011 &  & -0.027 &  & -0.073 & ** & -0.017 &  & -0.165 & *** & -0.020 &  & -0.195 & ***  \\
				\multicolumn{3}{l|}{~}  & (0.019 & ) & (0.025 & ) & (0.026 & ) & (0.025 & ) & (0.023 & ) & (0.024 & ) & (0.025 & ) & (0.026 & ) & (0.023 & ) & (0.026 & )  \\
				\multicolumn{3}{l|}{GICS sector indi.} & Incl. &  & Incl. &  & Incl. &  & Incl. &  & Incl. &  & Incl. &  & Incl. &  & Incl. &  & Incl. &  & Incl. &  \\
				\multicolumn{3}{l|}{Country of Risk indi.} & Incl. &  & Incl. &  & Incl. &  & Incl. &  & Incl. &  & Incl. &  & Incl. &  & Incl. &  & Incl. &  & Incl. &  \\
				\multicolumn{3}{l|}{Constant} & 7.284 & *** & 7.538 & *** & 7.662 & *** & 7.404 & *** & 5.986 & *** & 6.836 & *** & 6.790 & *** & 8.013 & *** & 5.599 & *** & 8.003 & *** \\
				\multicolumn{3}{l|}{~} & (0.159 & ) & (0.201 & ) & (0.207 & ) & (0.194 & ) & (0.267 & ) & (0.285 & ) & (0.271 & ) & (0.314 & ) & (0.303 & ) & (0.350 & ) \\
				\hline
				\multicolumn{3}{l|}{Observations} & 676 &  & 676 &  & 676 &  & 676 &  & 676 &  & 629 &  & 676 &  & 462 &  & 676 &  & 449 & \\
				\multicolumn{3}{l|}{Dropped obs. (M1)} & N/A &  & 0 &  & 0 &  & 0 &  & 0 &  & 47 &  & 0 &  & 214 &  & 0 &  & 227 &  \\
				\multicolumn{3}{l|}{$R^2$} & 0.6806 &  & 0.6997 &  & 0.6994 &  & 0.6861 &  & 0.7288 &  & 0.7568 &  & 0.7072 &  & 0.8308 &  & 0.7319 &  & 0.8395 & \\
				\multicolumn{3}{l|}{Adjusted R2} & 0.6642 &  & 0.6832 &  & 0.6830 &  & 0.6690 &  & 0.7136 &  & 0.7425 &  & 0.6907 &  & 0.8190 &  & 0.7163 &  & 0.8275 & \\
				\multicolumn{3}{l|}{$\Delta R^2$ (bps; M1)} & N/A &  & 191 &  & 188 &  & 55 &  & 482 &  & 762 &  & 266 &  & 1502 &  & 513 &  & 1589 &   \\
				\hline
			\end{tabular}
		}
		\caption{Standard deviations in parentheses. $^*p<0.05; ^{**}p<0.01; ^{***}p<0.001$}
		\label{tab:resultsIntermediateB}
	\end{tiny}
\end{table}

Tables~\ref{tab:resultsIntermediate} and \ref{tab:resultsIntermediateB} shows the intermediate models. Model 1 is the control model in Table~\ref{tab:results}. Models 2-11 are each of the concentration variables added to the control model. As they simply tests suppliers \textit{or} customers, none of these models are in Table~\ref{tab:results}. Models 12-14 show the results from suppliers and customers-pairs of the concentration variables. Model 12 was selected due to the simpler variables and the largest model improvement (i.e., the concentration model; model 2 in Table~\ref{tab:results}). Models 15-18 show the result of the concentration model with each of the influence variables added separately. It is worth noting that observations are dropped in model 16 and 18 due to missing data. Model 19 and 20 show the results when suppliers and customers-pairs of the simple average and weighted average variables were added, respectively. Model 20 drops 227 observations as missing data is present in the weighted average variables. We selected model 19 to be the full model due to the inclusion all observation even though the $R^2$ is 1,076bps better in model 20.

\end{document}